\documentclass[sigconf]{acmart}
\usepackage{amsmath}
\usepackage{booktabs} 
\usepackage[algo2e,lined,ruled,linesnumbered]{algorithm2e}
\usepackage{hyperref}
\usepackage[roman]{parnotes}
\setcopyright{rightsretained}

\acmDOI{10.475/123_4}

\acmISBN{123-4567-24-567/08/06}

\acmConference[K-CAP 2017]{ACM K-CAP conference}{Dec 2017}{Austin, Texas USA} 
\acmYear{2017}
\copyrightyear{2017}

\acmPrice{15.00}

\begin{document}
\title{Enriching Linked Datasets with New Object Properties}

\author{Subhashree S}
\affiliation{%
  \institution{Department of Computer Science and Engineering,\\
Indian Institute of Technology - Madras
}
  \city{Chennai, India} 
}
\email{ssshree@cse.iitm.ac.in}

\author{P Sreenivasa Kumar}
\affiliation{%
  \institution{Department of Computer Science and Engineering,\\
Indian Institute of Technology - Madras
}
  \city{Chennai, India} 
}
\email{psk@cse.iitm.ac.in}

\renewcommand{\shortauthors}{Subhashree S and P Sreenivasa Kumar}

\begin{abstract}
Although several RDF knowledge bases are available through the LOD initiative, the ontology schema of such linked datasets is not very rich. In particular, they lack object properties. The problem of finding new object properties (and their instances) between any two given classes has not been investigated in detail in the context of Linked Data. In this paper, we present DART (\textbf{D}etecting \textbf{A}rbitrary \textbf{R}elations for enriching \textbf{T}-Boxes of Linked Data) - an unsupervised solution to enrich the LOD cloud with new object properties between two given classes. DART exploits contextual similarity to identify text patterns from the web corpus that can potentially represent relations between individuals. These text patterns are then clustered by means of paraphrase detection to capture the object properties between the two given LOD classes. DART also performs fully automated mapping of the discovered relations to the properties in the linked dataset. This serves many purposes such as identification of completely new relations, elimination of irrelevant relations, and generation of prospective property axioms. We have empirically evaluated our approach on several pairs of classes and found that the system can indeed be used for enriching the linked datasets with new object properties and their instances. We compared DART with newOntExt system which is an offshoot of the NELL (Never-Ending Language Learning) effort. Our experiments reveal that DART gives better results than newOntExt with respect to both the correctness, as well as the number of relations.
\end{abstract}

\keywords{Linked Data, LOD enrichment, arbitrary relations, object properties, grounding}

\maketitle

\section{Introduction}\label{intro}
The Linked Data initiative provides a set of guidelines and best practices for publishing structured data and representing attribute values and relations among a set of entities.  The Linking Open Data (LOD) community project\footnote{\url{http://www.w3.org/wiki/SweoIG/TaskForces/CommunityProjects/LinkingOpenData}} works with the main objective of publishing open datasets as RDF triples and establishing RDF links between entities (aka objects) from different datasets. LOD complements the world wide web with a data space of entities connected to one another with labelled edges, which represent the relations among entity pairs (or entities and literal values). Many organizations have built systems to exploit the power of Linked Data for specific purposes. For example, the British Broadcasting Corporation (BBC) uses linked datasets such as DBpedia \cite{dbpedia} to enable cross-domain navigation and enhanced search\footnote{\url{https://www.w3.org/2001/sw/sweo/public/UseCases/BBC/}} in their websites. IBM has been using Linked Data as an integration technology for several years and their new cognitive system, Watson, has DBpedia and YAGO \cite{yago} as part of its major data sources~\cite{watson}.  

Currently, most linked datasets are rich in A-Box assertions but poor in T-Box information i.e they have a very weak ontology schema. They especially lack object properties. For example, the linked dataset YAGO has 488,469 classes \cite{yago3}. Among such a huge number of classes, surprisingly there are only 32 object properties\footnote{\url{http://www.mpi-inf.mpg.de/departments/databases-and-information-systems/research/yago-naga/yago/statistics/} - totally there are 60 object properties, but 28 of them connect the domain class to the class http://dbpedia.org/class/yago/YagoLiteral} and hence looking for more object properties to connect these classes becomes an interesting task. Adding more object properties to the ontology schema will help in enriching the domain being represented in the linked dataset. Question answering systems can make use of these additional relations to answer more number and also a wider range of questions.  To realize the full potential of Linked Data in various applications, it is important to enrich LOD with as many appropriate ontological axioms and assertions as possible.

Most of the Linked Data enrichment works (surveyed in \cite{swjsurvey}) focus on adding more instances to existing object properties (in this paper, the term `relation' is used as a synonym of `object property'). There are not many techniques
available in the literature that identify new relations, given two LOD classes.

The systems proposed in (\cite{ontext}, \cite{newontext}) for the purpose of extending the NELL ontology, OntExt and newOntExt respectively, can be adapted to the Linked Data settings to discover new object properties between given LOD classes. However, we found the following issues with their working: newOntExt tends to miss out important relations. It seeks to represent relations as text patterns and cluster patterns based on how frequently they co-occur with a pair of entities in a text corpus. The system tends to group dissimilar patterns into the same cluster and finally only the representative relation of the cluster is output by the system as a newly discovered relation. For example, given the classes athletes and sportsleagues as inputs, newOntExt places the relations ``doesn't play at" (currently not playing) and ``wants to play at" (wish to play) in the same cluster \cite{disslfn} because these two relations occur between the same subject-object pairs with a high frequency. Hence, only one of them gets selected as the cluster's representative relation, though both of them are correct relations, but with different meanings. Also, newOntExt does not do any contextual check to see if the pattern actually fits the context of the given two classes. For example, between the classes Languages and Countries, an incorrect pattern ``are people living in" is obtained from the web corpus as ``Chinese" can refer to both the language as well as the ethnic group. newOntExt does not perform any contextual check to eliminate such a pattern. 

In this paper, we present DART (\textbf{D}etecting \textbf{A}rbitrary \textbf{R}elations for enriching \textbf{T}-Boxes of Linked Data) which adopts an unsupervised approach in order to discover and add new object properties and their instances to a linked dataset. DART exploits contextual similarity tools and paraphrase detection in order to identify the correct set of text patterns which are most-likely to be useful as object properties between the two given LOD classes.  Additionally, it \emph{grounds} the relations to the linked dataset in order to identify the completely new relations and is also capable of generating candidate property axioms. By grounding, we mean mapping of discovered relations to existing LOD object properties.

To summarize, our contributions include the following:
\begin{enumerate}
\item{Given two classes belonging to a linked dataset, the proposed system DART discovers relations between them by exploiting text patterns from the web corpus, hence enriching the T-Box of the linked dataset. For example, given  the two classes, Religions\footnote{\url{http://dbpedia.org/class/yago/Religion105946687}\label{relig}} and Countries\footnote{\url{http://dbpedia.org/class/yago/Country108544813}\label{cntry}}, DART generates relations such as ``became the official religion in", ``is the predominant religion in" etc. }
\item{For each generated relation, a set of paraphrases are also generated that can be used to extract additional instances of the relation.}
\item{DART produces instances of the newly generated relations, leading to the enrichment of the A-Box. Continuing with the above example, it can add triples of the form (Hinduism, \emph{became the official religion in}, Nepal), (Christianity, \emph{is the predominant religion in}, Australia) etc.}
\item{A completely automated technique for grounding of the generated relations in the linked dataset has been proposed which also suggests T-Box axioms for the newly generated relations. For example, in the case of Empires\footnote{\url{http://dbpedia.org/class/yago/Empire108557482}\label{emp}} and Rulers\footnote{\url{http://dbpedia.org/class/yago/Ruler110541229}\label{rul}}, DART infers that the newly generated relation ``was ruler of" might be a sub-property of the YAGO property ``isLeaderOf". }
\item{Through the process of grounding, DART also eliminates irrelevant and ambiguous relations. 
}
\end{enumerate}
Our experiments show that DART gives much better results than newOntExt in terms of both precision and recall on input classes belonging to different domains. DART is also capable of suggesting insightful property axioms. \\
The rest of the paper is structured as follows: Section \ref{relworks} describes the related works from the literature. Section \ref{working} gives an account of the working of DART with each phase of the approach explained in detail. The experiments conducted by us in order to evaluate the effectiveness of the approach are presented in Section \ref{exp} along with the comparison of DART with newOntExt. Conclusions drawn from the work are given in Section \ref{concl}.

\section{Related Works}\label{relworks}
Relation enrichment (of those other than the \emph{owl:sameAs} links) of the linked datasets for the purpose of the overall growth of the LOD cloud has been the major focus in many recent works (surveyed in \cite{swjsurvey}). Most of the relation enrichment approaches surveyed in \cite{swjsurvey} focus on extracting more instances (subject-object pairs) of existing relations in the linked datasets. Works such as (\cite{triplifying1}, \cite{triplifying2}), \cite{T2Kmatch}, (\cite{mulwad1}, \cite{t2ld1}, \cite{t2ld2} and \cite{mulwad2}) and \cite{iitb} use the technique of interpreting web tables for this purpose and a few other works such as \cite{sofie} and  \cite{distantlargescale} propose using various semi-supervised approaches for the same. Distant supervision is another new paradigm which has been recently adopted by many works (\cite{dist}, \cite{mintz}, \cite{dbpediadist},  \cite{classhierarchy}, \cite{yagowikipedia}) in order to extract more instances of existing relations. Distant supervision is  the  technique  of  utilizing  a  large  number  of known facts (from a huge linked dataset such as Freebase) for automatically labeling  mentions  of  these  facts  in  an  unannotated text corpus, hence generating training data. A classifier is learnt based on this weakly labeled training data in order to classify unseen instances \cite{dist}.  

Apart from enriching the datasets with additional instances of already existing relations, two other less-explored problems of relation enrichment are: (1)finding instances of specified new relations and (2) discovering arbitrary new relations. 

By instances of specified new relations, we mean that the relation is not present in the dataset currently but the name of the relation is given to the system and the system needs to add instances of such a relation to the dataset. The technique proposed in \cite{plato} to detect instances of ``part-of'' (partonomy) relation between linked data instances falls under this category. Similarly, the SILK link discovery framework \cite{silk} which is primarily used to detect \emph{owl:sameAs} links is also capable of detecting instances of user-specified relations. It uses its own declarative language, Silk - Link Specification Language (Silk-LSL) in order to specify the two datasets between which the links ought to be found and to give the link type. Coming to the second problem of discovering arbitrary new relations, it can be defined as the task of finding any or all possible relations between two given classes. We find that it has not been tackled by many works. It is precisely this problem we address in this paper. It is to be noted that the system is not aware of the possible relations between the concerned classes before-hand and hence such relations are termed as arbitrary relations (as defined in \cite{reverb}). 
There are two systems, OntExt \cite{ontext} and newOntExt (\cite{newontext}, \cite{newontextjournal}) which have been proposed in the context of helping NELL to extend its ontology by means of discovering new relations between the ontology classes. They are described below:

\textbf{OntExt:}
Given two noun categories (\cite{nell} calls classes as noun categories), and their instances, OntExt discovers relations between them by exploiting the notion that similar patterns occur between the same subject-object pairs. For example, if the patterns ``Ganges flows through Allahabad" and ``Ganges in the heart of Allahabad" occur in the web corpus with a very high frequency then this can be taken as an indicator that the patterns, ``flows through" and ``in the heart of" are similar to each other. When such an evidence is shown by many number of subject-object pairs, OntExt gives a very high similarity score between the two patterns. In general, OntExt works in the following manner: given a pair of categories  and a set of sentences-each containing a pair of instances known to belong to the given categories, OntExt collects the words in between the instances from each sentence and calls these words a ``context-pattern". Then it builds a context-pattern by context-pattern co-occurrence matrix based on the frequencies of occurrence of these context-patterns with the same subject-object instance pairs. For example, in the above case of finding relations between Rivers and Cities, if the pair ``Ganges" and ``Allahabad" occurs with the context-pattern  ``flows through" with a frequency $f_1$ and the pair occurs with the pattern ``in the heart of" with a frequency $f_2$, then the matrix entry corresponding to these two context-patterns will be given a value of $(f_1+f_2)$. In case there is another subject-object pair (for example- Thames, London) occurring with both these context-patterns with frequencies $f_3$ and $f_4$ respectively, then the matrix cell value becomes $(f_1+f_2+f_3+f_4)$. K-means clustering is applied on the normalized matrix to group the related context-patterns together. The centroid of each cluster is proposed as a new relation. Then the subject-object pairs are ranked based on how often they occur along with each context-pattern using the formula in equation (1). Finally, the top 50 subject-object pairs are given as seed instances of the new relation to NELL \cite{nell}.\\ 
Weight of a (subject,object) pair ``s"\\
\begin{equation}
\begin{aligned}
= \sum_{c \in cluster}\frac{Occ(c,s) }{1+ sd(c) }
\end{aligned}
\end{equation}
Where,\\
cluster is the cluster of pattern contexts for the given new relation, \\
Occ(c,s) is the number of times instance ``s" co-occurs with the context pattern ``c", \\
sd(c) is the standard deviation of the context pattern from the centroid of the pattern cluster\\

As more than half of the relations generated by OntExt were invalid (determined manually in \cite{ontext}), the authors of OntExt have proposed a classifier which can differentiate between valid and invalid relations to some extent. 

\textbf{newOntExt:}
newOntExt which was developed based on OntExt had a few changes in its working \cite{newontext}: instead of considering all the words in between the two input instances as a pattern, newOntExt used ReVerb \cite{reverb} for extracting the patterns in order to reduce the number of noisy patterns obtained; for optimising the computational cost, a more elegant file structure was used for searching through the sentences; instead of considering every pair of categories as input to this system, reduced category groups of interest were formed to pick the input category pairs. 

A major difference between DART and newOntExt is that the latter takes co-occurrence values of the patterns to be an indicator of the semantic similarity between them whereas DART computes the semantic similarity by means of paraphrase detection techniques. It should be noted that DART does not rely upon the lexical similarity of the patterns i.e DART can detect the semantic similarity even if the two patterns have disjoint set of words. In addition to this, DART also performs grounding of relations and generation of candidate property axioms. Comparison of DART with newOntExt is described in Section \ref{eval}.

\section{Working of DART}\label{working}
\subsection{Pre-processing}\label{preprocess}
Given two classes D1 and D2, we need patterns occurring in the web corpus along with the instances of D1 and D2, in order to discover the possible relations between them. Hence for this purpose, we obtain (subject, predicate, object) triples - known as a triple corpus C from the RCE 1.1 file  \footnote{ReVerb ClueWeb Extractions 1.1: dataset consisting of 15 million triples produced by running ReVerb on the English portion of ClueWeb09 corpus}, such that the subject and object belong to D1 (D2) and D2 (D1) respectively. We have used the RCE dataset in our experiments as newOntExt employs ReVerb and we wished to maintain uniform set of inputs for both DART and newOntExt for a fair comparison. However we can also replace this step in the following manner: use a web corpus such as ClueWeb and extract sentences containing instances of D1 and D2; then apply any triplification tool such as ClausIE \cite{clausie}, Ollie \cite{ollie} etc to obtain the input triples corpus C.

We also store the direction of these triples in C, i.e if the subject of the triple belongs to D1 and the object belongs to D2, then the direction is marked as ``forward". If subject belongs to D2 and object belongs to D1, the direction is marked as ``reverse". 
\subsection{Relation discovery phase}\label{rdphase}
Relation discovery phase, given in Algorithm \ref{algo:RelDisc}, takes the corpus C as input and outputs clusters of synonymous relations. We collect all the unique predicates in C (let us call them ``patterns") and filter them based on whether they are suitable for the given input domain or not (Lines 1-7) i.e a contextual similarity check is performed in the following manner: in each pattern, all the function words \footnote{\url{http://www.sequencepublishing.com/academic.html}} are removed (as they are not context-specific words) and the remaining words are checked for similarity with the domain name. For example, let us assume that the user intends to find the relations between a set of rivers and a set of cities, and the user-specified domain name is ``river". If the pattern under consideration is ``rises in", DART checks the similarity of ``rises" (as the other word ``in" is a functional word) with ``river" and if this similarity crosses a certain threshold (more details on how this threshold was fixed are given in Section \ref{exp}), DART includes this pattern else discards it. We use the Word2Vec \cite{word2vec} model proposed and trained by Google \footnote{\url{https://code.google.com/archive/p/word2vec/}} for finding the contextual similarity. The intuition behind this step is that, patterns not relevant to the domain obtained from the web corpus can be eliminated by checking if the contexts of the pattern and the domain name are close to each other, i.e this serves as a pseudo disambiguation step.  The filtered patterns are then subjected to single pass clustering \cite {singlepassclustering}. Single pass clustering works as follows (Lines 8-33): Take each pattern ``p" and check its semantic similarity with the representative relations of all the clusters. Place ``p" in the cluster whose representative relation has the maximum similarity with it. Now recompute the representative relation for this augmented cluster in the following manner - representative relation is the pattern which has the maximum average similarity with the other patterns in that cluster. If the maximum similarity value is lesser than a fixed threshold value (=0.5), place ``p" in a new cluster.

In order to determine the semantic similarity between two patterns, we modified the paraphrase detection technique proposed by Mihalcea et al. \cite{mihalcea}: We have eliminated the word specificity weights. In \cite{mihalcea}, the individual word-to-word similarity values were weighted using a word specificity measure so that higher importance can be given to a semantic matching identified between two specific words such as ``collie" and ``sheepdog" when compared to a matching identified between words such as ``get" and ``become". In the context of DART, words such as ``get" and ``become" (any verb in general) have a good chance of occurring in the input patterns as the aim of DART is to extract relations between classes. Hence, giving a low weight to such words (as done in \cite{mihalcea}) is not appropriate in the context of DART.
The formula used to determine similarity of patterns in our work is given in equation (2). 
\begin{equation}
\begin{aligned}
sim(T_1,T_2) = 1/2
(\frac{\sum_{w \in T_1} (maxSim(w,T_2)) }{len(T_1) } + \\
\frac{\sum_{w \in T_2} (maxSim(w,T_1)) }{len(T_2) })
\end{aligned}
\end{equation}

where,\\
$T_1$ and $T_2$ represent the input text segments,\\
$maxSim(w,T_i)$ refers to the similarity value of the word in $T_i$ which is most similar to the word $w$ in the other text segment,\\
$len(T_i)$ refers to the number of words in $T_i$\\

In our implementation, the threshold value chosen to consider two segments $T_1$ and $T_2$ to be similar is 0.5 (adopted from \cite{mihalcea}).

LESK \cite{lesk} has been used to perform the word-to-word similarity component of equation (2), as it works for all combinations of parts of speech. The representative relations of the clusters obtained at the end of this phase form the relations between the two given classes. 

\begin{algorithm2e}
\DontPrintSemicolon
\KwIn{$C$, the corpus;\\
~~~~~~~~~~~~$dname$, user-specified domain name; \\
~~~~~~~~~~~~$cThreshold$, the threshold for contextual similarity;\\
~~~~~~~~~~~~$sThreshold$, the threshold for semantic similarity;
}
\KwOut{$clusts$, clusters of relations;\\
~~~~~~~~~~~~$repRels$, representative relations; \\
}
$allPatterns \gets$ set of all patterns in $C$  \\
$filteredPatterns \gets \emptyset$ \\
\For {each pattern $p \in allPatterns$}{
	\If{any word in $p$ has wordToVec 		similarity with $dname$  >= $cThreshold$}
	{ add $p$ to $filteredPatterns$ }

}
$clusts \gets \emptyset$ \\
$repRels \gets \emptyset$ \\
\For {each pattern $p \in filteredPatterns$}{

$maxRel \gets null$ \\
\If{$repRels = \emptyset$}
	{add $p$ to a new cluster $cl$ \\
	add $p$ to $repRels$ \\
	add $cl$ to $clusts$ \\
	}
\Else{
	$maxRel \gets$ the representative relation which has the maximum similarity $maxSim$ with $p$\\
	\If{$maxSim$ >= $sThreshold$}
	{
	add $p$ to $cl1$, the cluster containing $maxRel$ ;\\
	\For {each pattern $p1 \in cl1$}{
	calculate $p1$'s average similarity with other patterns in $cl1$
	
	}
	$rep \gets$ pattern in $cl1$ having maximum average similarity\\
	add $rep$ to $repRels$ \\
	}
	\Else{
	add $p$ to a new cluster $cl$ \\
	add $p$ to $repRels$ \\
	add $cl$ to $clusts$ \\
	}
	}
	
}

\Return{$clusts$, $repRels$}\;
\caption{{\sc Relation-Discovery}}
\label{algo:RelDisc}
\end{algorithm2e}

\subsection{Grounding of relations}\label{grounding}
Once the relation discovery phase generates relations between the two input classes, the system needs to check if these relations can be grounded in the linked dataset, i.e whether they can be mapped to some existing property in the linked dataset. Only if a relation cannot be grounded (mapped to existing LOD properties), it is added as a new relation leading to T-Box enrichment of the linked dataset. In order to map a  representative relation with existing LOD properties, DART checks the semantic similarity between an LOD property $p$ and the representative relation $r$  (using equation (2), but with an increased threshold of 0.75 as we want to avoid spurious mappings). If the similarity value crosses 0.75, then the similarity between $p$ and every relation in the cluster of $r$ is determined. Finally, if more than 50\% of the relations in $r$'s cluster have a similarity value >= 0.75 with $p$, then $r$ is said to be grounded and matched to $p$.

In order to determine the domain and range of the LOD property to which the relation was grounded, we use the ontology of the linked dataset (if the ontology lacks this information, we use the system proposed in \cite{topper}). The domain and range of the grounded relation are the 2 input classes. Then these grounded relations are handled by DART in two ways: If the grounded relation $r$ and the matched LOD property $p$ have the same domain and range then it means we have detected a new equivalent property to $p$. If the domain of $r$ is the range of $p$ and if the range of $r$ happens to be the domain of $p$, then it means $r$ is a new inverse property for $p$. If the domain (and/or range) of $r$ is a subclass of the domain (and/or range) of $p$, then it means we have discovered a new sub-property for the LOD property $p$. Hence in these cases, we don't discard the relation completely. More relation instances of $r$ are produced by DART in the Triple-Finding phase (Section \ref{tfphase}). However if the domain and range do not match in any of the above mentioned ways, then we consider the grounded relation ``r" as an \emph{ambiguous} or irrelevant (noisy) relation and hence completely discard it. In this paper, we call a relation ambiguous if it holds between 2 or more pairs of classes. For example, the relation ``caused by" is an ambiguous relation as it is a meaningful relation between the classes (Event, Event) as well as between the classes (Disease, Drug).

Note that there are a few other works in the literature (such as \cite{farber}) which focus mainly on the grounding of relations in a Knowledge Base (KB). However, the goal of such systems and the goal of DART are very different from each other with respect to grounding - the former kind of systems extract triples from an external source (such as text) and attempt to ground the relations.  They retain only the grounded relations and consider the non-grounded relations as irrelevant to the schema and hence discard them. In our system, we attempt to ground the discovered relations in a linked dataset in order to achieve three things:  identify irrelevant relations among the  grounded relations and discard them; align the remaining grounded relations to the ontology schema and generate prospective axioms such as inverse, subproperty etc.; identify the non-grounded (new) relations and add them to the T-Box. Moreover in systems such as \cite{farber}, grounding of relations is based on the grounding of entities. We do not use this approach as it will not help us to identify irrelevant and ambiguous relations. Also, the method in \cite{farber} is semi-automatic i.e a human is involved  to decide whether ``buy" can be mapped to the relation ``acquired". On the contrary, DART performs this phase in an automated fashion.

\subsection{Triple finding phase}\label{tfphase}
In this phase, we intend to find all triples (s,p,o) where p is a relation found in the previous phase, hence enriching the A-Box of the linked dataset. In order to obtain the instances of a new relation (let us call this ``p"), each relation ``r" in p's cluster is looked up in the corpus C, and the subject-object pair found in C for the relation ``r"  is given as an instance to ``p". Hence $(subject, p, object)$ becomes the final triple. One thing to be noted here is if the relation looked up in the corpus (``r") is of forward direction and the relation ``p" is of reverse direction (notion of directions explained in the Section \ref{preprocess}), then the final triple given as output becomes $(object, p, subject)$.

\section{Experiments and Results}\label{exp}
The proposed system, DART, has been implemented in Java 1.7 and all experiments have been conducted on a Linux system equipped with an Intel 3.20 GHz quad-core processor and 32 GB main memory. All details regarding the input classes, the relations and relation instances obtained can be found in our project web page\footnote{\url{https://sites.google.com/site/ontoworks/projects}\label{wp}}. 

The experiments conducted on the NELL Knowledge Base for the purpose of comparing DART with newOntExt, and the observations made are given in Section \ref{eval}.  The details about the experiments held to gauge DART's performance on LOD classes are given in Section \ref{appl}.

\subsection{Comparison with newOntExt}\label{eval}
Though the primary aim of DART is to enrich linked datasets such as DBpedia, YAGO etc., in this subsection, we have conducted experiments on collections of entities belonging to the NELL Knowledge-Base\footnote{NELL.08m.1050.esv.csv ``every belief in the KB" file downloaded from \url{http://rtw.ml.cmu.edu/rtw/resources} on 26th April 2017} in order to compare DART against the newOntExt system. We have used the implementation of newOntExt provided by the authors of \cite{newontext}\footnote{\url{https://github.com/MaLL-UFSCar/ontext}}. The systems have been compared using two measures, accuracy and the number of meaningful (the terms ``meaningful" and ``correct" have been used synonymously in the paper) relations obtained. Accuracy is taken as the ratio of the correct relations (as determined by human evaluators) to the total number of relations obtained.  

For newOntExt, the value of k used in the k-means clustering of patterns (see Section \ref{relworks}) affects the quality of the relations obtained to a large extent. Since the value of k used is not mentioned in (\cite{newontext}, \cite{newontextjournal}) and has been fixed in a dataset-specific manner in \cite{ontext}, we have applied the Elbow method \cite{mmdbook} to determine the best k value (from a range of k=3 to k=29) for clustering the patterns, for each experiment. For DART, the threshold used for checking the contextual similarity using Word2Vec has an impact on the quality of the relations obtained. Hence we conducted experiments for each input class pair for 5 different thresholds - 0.1, 0.2, 0.3, 0.5 and 0.7. We observed that the thresholds of 0.3, 0.5 and 0.7 give very meaningful but very less number of relations. On the other hand, setting the threshold to 0.1 gives very high number of relations (around 130) but most of them are noisy, irrelevant relations. Therefore we decided to use the threshold of 0.2 uniformly for our experiments in order to maintain a good trade-off between the correctness and the number of relations obtained (however, the user can choose to vary this threshold depending on the requirements of the application). For the evaluation, the relations were presented in this format: <classname> relation <classname>(for example, <rivers> flows through <cities>) and three ontology engineers were assigned to evaluate them on a two-valued scale: \emph{correct}, and \emph{incorrect}. We required that all the three evaluators agree that a relation is correct in order for it to be counted as correct. Table \ref{tab:sample} gives details about the input categories and a few sample relations obtained through DART. We have chosen the input categories such that they belong to different domains (Geography, Industries and Medicine) in order to demonstrate the versatility of the proposed system. Also, these particular categories were chosen from their respective domains to ease the process of manual evaluation. Table \ref{tab:accuracy} gives the accuracy and the number of correct relations obtained through DART and newOntExt. 

\begin{table*}
  \caption{Input categories from the NELL KB and sample relations through DART}
  \label{tab:sample}
  \begin{tabular}{ccl}
    \toprule
    \parnoteclear 
    D1 (size) & D2 (size) & Sample relations through DART\\
    \midrule
    Rivers\parnote{\url{http://rtw.ml.cmu.edu/rtw/kbbrowser/pred:river}} (21059) & Cities\parnote{\url{http://rtw.ml.cmu.edu/rtw/kbbrowser/pred:city}} (26119) &  ``flows through", ``is just a few miles west of", ``drowned in" \\
   Languages\protect\parnote{\url{http://rtw.ml.cmu.edu/rtw/kbbrowser/pred:language}}  (11278) &  Countries\parnote{\url{http://rtw.ml.cmu.edu/rtw/kbbrowser/pred:country}}  (3064) & ``are spoken in", ``is a common language in", ``is an official language in"\\
   Vegetables\parnote{\url{http://rtw.ml.cmu.edu/rtw/kbbrowser/pred:vegetable}}  (258) & Diseases\parnote{\url{http://rtw.ml.cmu.edu/rtw/kbbrowser/pred:disease}}  (16120) & ``is good for curing", ``increases the risk for"\\
   CEOs\parnote{\url{http://rtw.ml.cmu.edu/rtw/kbbrowser/pred:ceo}}  (7289) & Companies\parnote{\url{http://rtw.ml.cmu.edu/rtw/kbbrowser/pred:company}}  (41660) & ``is ceo of", ``is a founder of", ``is a company established by" \\
    \bottomrule
  \end{tabular}
  \parnotes
 
\end{table*}

\begin{table*}
  \caption{Evaluation Results-accuracy and number of meaningful relations obtained}
  \label{tab:accuracy}
  \begin{tabular}{ccccc}
    \toprule
     Input categories & DART & &     newOntExt (with best k-value) &  \\
    \midrule
      & no. of correct relations & accuracy & no. of correct relations & accuracy\\
    \midrule
Rivers, Cities & 15 & 0.42 & 4 & 0.15 \\
Languages, Countries & 22 & 0.63 &  7 & 0.54\\
Vegetables, Diseases & 22 & 0.88 &  16 & 0.84\\
CEOs, Companies &  19 & 0.86 & 11 & 0.58\\
    \bottomrule
  \end{tabular}
\end{table*}

From Table \ref{tab:accuracy} we can see that DART performs better than newOntExt both as a recall-oriented system and as a precision-oriented system. Since clustering of patterns in newOntExt is based on co-occurrence values, dissimilar(but meaningful) patterns tend to get grouped together and hence many meaningful patterns get lost, leading to lower number of correct relations from newOntExt. For example, in the experiment conducted on the classes CEO and Company, newOntExt places the patterns ``is the ceo of" and ``is the founder of" into the same cluster because the two patterns occur between the same set of subject-object pairs. Only one pattern from a cluster gets chosen as the centroid of the cluster and output by newOntExt and hence the other pattern is dropped though it is a meaningful relation between the given classes. Also, the Word2Vec model used by DART has eliminated irrelevant patterns such as ``are people living in" (in the case of Languages and Countries) leading to a better accuracy value of DART. 

\subsubsection{Grounding in the context of NELL relations}
The convention followed by NELL and the LOD for naming the relations are different. In NELL, the domain and/or range names are appended to the actual relation to form the relation name. For example, the relation ``flows through" which holds between the classes Rivers\footnote{\url{http://rtw.ml.cmu.edu/rtw/kbbrowser/pred:river}\label{rv}} and Cities\footnote{\url{http://rtw.ml.cmu.edu/rtw/kbbrowser/pred:city}\label{ct}} is named ``riverflowsthroughcity" (in LOD, such a relation would be named ``flowsThrough"). Similarly, the relation ``side effect caused by" which holds between the classes Physiological Condition and Drugs is named ``sideeffectcausedbydrug" in NELL. The advantage of using such a naming technique is that every sense of the relation can be captured through its name itself, thus giving no room for ambiguity. Hence there is no necessity for grounding the generated relations in the NELL Knowledge Base. Also, the main goal of DART is to enrich the LOD and hence we exclude the process of grounding in our experiments on the NELL KB. We compare the number of correct relations obtained through DART and those obtained through newOntExt (see Table \ref{tab:accuracy}) irrespective of whether they are already present in the NELL KB. This is to demonstrate the efficacy of DART vs newOntExt in the context of discovering relations between given classes.

\subsubsection{Complexity of DART vs newOntExt}
Table \ref{tab:time} gives the details about the time taken by DART and newOntExt for the four experiments. 
\begin{table}
  \caption{Time taken (in seconds)}
  \label{tab:time}
  \begin{tabular}{ccc}
    \toprule
     Input classes & DART & newOntExt  \\
    \midrule
      
Rivers, Cities & 68 & 22.98\\
Languages, Countries & 429 & 9.5 \\
Vegetables, Diseases & 37 & 6.51 \\
CEOs, Companies & 5 & 6.23 \\
    \bottomrule
  \end{tabular}
\end{table}

As newOntExt follows co-occurrence based clustering of patterns and DART performs semantic similarity check for clustering of patterns, the time taken by DART would be inherently higher than newOntExt. However we have attempted to reduce the computational complexity in two ways: by employing Word2Vec to filter patterns and by using single-pass clustering to cluster the patterns(as opposed to clustering algorithms like k-means which perform several iterations). For example, in the case of CEOs and Companies the initial number of patterns was 339, whereas after filtering through Word2Vec the number of patterns remarkably reduced to 51. Hence the final number of patterns subjected to clustering is low leading to a reduced consumption of time (even lesser than newOntExt). In most of the cases DART takes only around few seconds to 1 minute to perform its task (except for the case of Languages and Countries where 230 patterns are output by the Word2Vec stage and subjected to clustering). It is an interesting piece of future work to further optimize the working of DART.

\subsection{Evaluation of DART on linked datasets}\label{appl}
In this Section, we give an account of the experiments held to demonstrate the enrichment of LOD through DART, i.e we have chosen classes from linked datasets such as YAGO and DBpedia as our input classes. Table \ref{tab:sample1} gives the details of the input classes taken and a few sample relations obtained through DART. Here again, we have chosen these classes from different domains (Geography, Literature, History and Music) to prove that our approach is versatile.

\begin{table*}
  \caption{Input classes from the LOD and sample relations through DART}
  \label{tab:sample1}
  \begin{tabular}{ccl}
    \toprule
    \parnoteclear 
    D1 (size) & D2 (size) & Sample relations through DART\\
    \midrule
    Religions\parnote{\url{http://dbpedia.org/class/yago/Religion105946687}} (222) & Countries\parnote{\url{http://dbpedia.org/class/yago/Country108544813}} (5726) &  ``became the official religion in", ``is the predominant religion in", ``is the fastest growing religion in" \\
   Empires\parnote{\url{http://dbpedia.org/class/yago/Empire108557482}} (325) &  Rulers\parnote{\url{http://dbpedia.org/class/yago/Ruler110541229}} (9118) & ``ascended the throne of", ``declared war on", ``inherited the kingdom of", ``is founded by" \\
   Writers\parnote{\url{http://dbpedia.org/class/yago/Writer110794014}} (10000) &  Novels\parnote{\url{http://dbpedia.org/class/yago/Novel106367879}} (10000) & ``is written by", ``is a novel by", ``is a biography of", ``is the award winning author of"\\
   Music genres\parnote{\url{http://dbpedia.org/ontology/MusicGenre}} (1245) & Music genres & ``is a subgenre of", ``is more popular than" \\
   \bottomrule
  \end{tabular}
   \parnotes
\end{table*}

Table \ref{tab:groundres} gives an account of few relations which were mapped to the LOD properties in each experiment, and the action performed by DART on the grounded relations. The full list of all the grounded relations is available in our project web page\textsuperscript{\ref{wp}}.
\begin{table*}
  \caption{Grounding of relations}
  \label{tab:groundres}
  \begin{tabular}{cccc}
    \toprule
    Input Classes & LOD property & Grounded relation & Action taken by DART  \\
    \midrule
      
Religions, Countries & isLeaderOf  & is the father of & Domain, range not matching-discard\\
\midrule
Rulers, Empires & isLeaderOf & was ruler of & Domain, range matched through subclass - candidate sub-property\\
\midrule
Writers, Novels & directed & directed by & Domain, range not matching-discard \\
\midrule
Music genres, music genres & musicSubgenre & is a subgenre of & Domain, range match - candidate equivalent property\\
    \bottomrule
  \end{tabular}
\end{table*}

Table \ref{tab:results} shows the accuracy and the number of correct relations obtained for the input classes in Table \ref{tab:sample1}. As done in Section \ref{eval}, three ontology engineers were asked to evaluate the relations manually and a relation was considered correct only if all the three experts agreed that it is correct.

\begin{table}
  \caption{Evaluation Results-Accuracy and number of meaningful relations obtained}
  \label{tab:results}
  \begin{tabular}{ccc}
    \toprule
    Input Classes & No. of correct relations & Accuracy  \\
    \midrule
      
Religions, Countries & 25 & 0.50   \\
Empires, Rulers & 10 & 0.833 \\
Writers, Novels & 15 & 0.52  \\
Music genres, Music genres & 9 & 0.69   \\
    \bottomrule
  \end{tabular}
\end{table}

\subsubsection{Value of grounding}
Following the grounding technique explained in Section \ref{grounding}, DART discarded or retained the grounded relations appropriately. It should be noted that if the discarded irrelevant relations (such as ``is the father of" in the case of Religions and Countries) had been included in the output of DART, then the accuracy of DART would have decreased. Hence, the grounding phase improves the performance of DART. The grounding phase also suggests candidate equivalence, sub-property and inverse property axioms between the relations and existing LOD properties. These property axioms can further be validated through techniques that are based on determining the support from the instances \cite{propaxms} and then added to the T-Box. We intend to do the validation and enrichment process as a part of our future work.
DART has not been compared with any of the property alignment systems (such as those surveyed in \cite{all3}) since the main goal of DART is to generate relations between two given classes only. DART suggests candidate property axioms which are yet to be supported by evidence from the A-Box. In that sense, DART can also be seen as a system which is capable of extracting new prospective inverse relations from text. For example, if one needs to find the inverse of  the DBpedia property ``author", the domain and range of ``author", namely the classes Person and Book can be given as inputs to DART and DART would produce the relations both in the forward direction (the same direction as ``author") as well as the reverse. If any of the relations in the reverse direction get grounded to the property ``author" (i.e the relation's direction is opposite to that of ``author" but its meaning is similar to ``author"), then that relation is a prospective inverse property to the ``author" property.

\section{Conclusions and Future Work}\label{concl}
The central idea behind this paper is to propose a completely automated and unsupervised technique to identify possible arbitrary relations between two classes of Linked Data. For this purpose, we have built a system, DART, whose working connects the techniques of contextual similarity checking and paraphrase detection into a unified framework for discovering new relations from the web patterns. DART then attempts to ground the discovered relations in the linked dataset in order to discard irrelevant relations and identify new relations. The fully automated grounding technique proposed in this paper also generates prospective property axioms for the enrichment of the linked dataset. 

The results gathered reveal the potential of DART to unearth many interesting relations between a given pair of classes thus leading to the growth of a relationship-rich LOD. DART outperforms the state-of-the-art system with respect to the validity as well as the number of relations. As a part of our future work, we intend to validate the grounding phase to improve its accuracy and efficiency. We would also like to propose methods to validate the prospective property axioms generated through DART by means of gathering evidence from the generated relation instances.

\bibliographystyle{ACM-Reference-Format}
\bibliography{ref} 


\begin{thebibliography}{00}


\ifx \showCODEN    \undefined \def \showCODEN     #1{\unskip}     \fi
\ifx \showDOI      \undefined \def \showDOI       #1{#1}\fi
\ifx \showISBNx    \undefined \def \showISBNx     #1{\unskip}     \fi
\ifx \showISBNxiii \undefined \def \showISBNxiii  #1{\unskip}     \fi
\ifx \showISSN     \undefined \def \showISSN      #1{\unskip}     \fi
\ifx \showLCCN     \undefined \def \showLCCN      #1{\unskip}     \fi
\ifx \shownote     \undefined \def \shownote      #1{#1}          \fi
\ifx \showarticletitle \undefined \def \showarticletitle #1{#1}   \fi
\ifx \showURL      \undefined \def \showURL       {\relax}        \fi
\providecommand\bibfield[2]{#2}
\providecommand\bibinfo[2]{#2}
\providecommand\natexlab[1]{#1}
\providecommand\showeprint[2][]{arXiv:#2}

\bibitem[\protect\citeauthoryear{Aprosio, Giuliano, and Lavelli}{Aprosio
  et~al\mbox{.}}{2013}]%
        {dbpediadist}
\bibfield{author}{\bibinfo{person}{Alessio~Palmero Aprosio},
  \bibinfo{person}{Claudio Giuliano}, {and} \bibinfo{person}{Alberto Lavelli}.}
  \bibinfo{year}{2013}\natexlab{}.
\newblock \showarticletitle{Extending the Coverage of DBpedia Properties using
  Distant Supervision over Wikipedia.}. In \bibinfo{booktitle}{{\em
  NLP-DBPEDIA@ISWC}} {\em (\bibinfo{series}{CEUR Workshop Proceedings})},
  \bibfield{editor}{\bibinfo{person}{Sebastian Hellmann},
  \bibinfo{person}{Agata Filipowska}, \bibinfo{person}{Caroline Barriere},
  \bibinfo{person}{Pablo~N. Mendes}, {and} \bibinfo{person}{Dimitris
  Kontokostas}} (Eds.), Vol.~\bibinfo{volume}{1064}.
  \bibinfo{publisher}{CEUR-WS.org}.
\newblock


\bibitem[\protect\citeauthoryear{Assis and Casanova}{Assis and
  Casanova}{2014}]%
        {classhierarchy}
\bibfield{author}{\bibinfo{person}{PedroH.R. Assis} {and}
  \bibinfo{person}{MarcoA. Casanova}.} \bibinfo{year}{2014}\natexlab{}.
\newblock \showarticletitle{Distant Supervision for Relation Extraction Using
  Ontology Class Hierarchy-Based Features}.
\newblock In \bibinfo{booktitle}{{\em The Semantic Web: ESWC 2014 Satellite
  Events}}, \bibfield{editor}{\bibinfo{person}{Valentina Presutti},
  \bibinfo{person}{Eva Blomqvist}, \bibinfo{person}{Raphael Troncy},
  \bibinfo{person}{Harald Sack}, \bibinfo{person}{Ioannis Papadakis}, {and}
  \bibinfo{person}{Anna Tordai}} (Eds.). \bibinfo{series}{Lecture Notes in
  Computer Science}, Vol.~\bibinfo{volume}{8798}. \bibinfo{publisher}{Springer
  International Publishing}, \bibinfo{pages}{467--471}.
\newblock
\showISBNx{978-3-319-11954-0}


\bibitem[\protect\citeauthoryear{Banerjee and Pedersen}{Banerjee and
  Pedersen}{2002}]%
        {lesk}
\bibfield{author}{\bibinfo{person}{Satanjeev Banerjee} {and}
  \bibinfo{person}{Ted Pedersen}.} \bibinfo{year}{2002}\natexlab{}.
\newblock \showarticletitle{An Adapted Lesk Algorithm for Word Sense
  Disambiguation Using WordNet}. In \bibinfo{booktitle}{{\em Proceedings of the
  Third International Conference on Computational Linguistics and Intelligent
  Text Processing}} {\em (\bibinfo{series}{CICLing '02})}.
  \bibinfo{publisher}{Springer-Verlag}, \bibinfo{address}{London, UK, UK},
  \bibinfo{pages}{136--145}.
\newblock
\showISBNx{3-540-43219-1}


\bibitem[\protect\citeauthoryear{Barchi and Hruschka}{Barchi and
  Hruschka}{2014}]%
        {newontext}
\bibfield{author}{\bibinfo{person}{P.~H. Barchi} {and}
  \bibinfo{person}{E.~Rafael Hruschka}.} \bibinfo{year}{2014}\natexlab{}.
\newblock \showarticletitle{Never-ending ontology extension through machine
  reading}. In \bibinfo{booktitle}{{\em 2014 14th International Conference on
  Hybrid Intelligent Systems}}. \bibinfo{pages}{266--272}.
\newblock


\bibitem[\protect\citeauthoryear{Barchi and Hruschka}{Barchi and
  Hruschka}{2015}]%
        {newontextjournal}
\bibfield{author}{\bibinfo{person}{P.~H. Barchi} {and}
  \bibinfo{person}{E.~Rafael Hruschka}.} \bibinfo{year}{2015}\natexlab{}.
\newblock \showarticletitle{Two different approaches to Ontology Extension
  Through Machine Reading}.
\newblock \bibinfo{journal}{{\em Journal of Network and Innovative
  Computing\/}} \bibinfo{volume}{3}, \bibinfo{number}{1}
  (\bibinfo{year}{2015}), \bibinfo{pages}{78--87}.
\newblock
\showISSN{2160-2174}


\bibitem[\protect\citeauthoryear{Bizer, Volz, Kobilarov, and Gaedke}{Bizer
  et~al\mbox{.}}{2009}]%
        {silk}
\bibfield{author}{\bibinfo{person}{Christian Bizer}, \bibinfo{person}{Julius
  Volz}, \bibinfo{person}{Georgi Kobilarov}, {and} \bibinfo{person}{Martin
  Gaedke}.} \bibinfo{year}{2009}\natexlab{}.
\newblock \showarticletitle{Silk - A Link Discovery Framework for the Web of
  Data}. In \bibinfo{booktitle}{{\em 18th International World Wide Web
  Conference}}.
\newblock


\bibitem[\protect\citeauthoryear{Carlson, Betteridge, Kisiel, Settles, Jr., and
  Mitchell}{Carlson et~al\mbox{.}}{2010}]%
        {nell}
\bibfield{author}{\bibinfo{person}{Andrew Carlson}, \bibinfo{person}{Justin
  Betteridge}, \bibinfo{person}{Bryan Kisiel}, \bibinfo{person}{Burr Settles},
  \bibinfo{person}{Estevam R.~Hruschka Jr.}, {and} \bibinfo{person}{Tom~M.
  Mitchell}.} \bibinfo{year}{2010}\natexlab{}.
\newblock \showarticletitle{Toward an Architecture for Never-Ending Language
  Learning}. In \bibinfo{booktitle}{{\em AAAI}},
  \bibfield{editor}{\bibinfo{person}{Maria Fox} {and} \bibinfo{person}{David
  Poole}} (Eds.). \bibinfo{publisher}{AAAI Press}.
\newblock


\bibitem[\protect\citeauthoryear{Del~Corro and Gemulla}{Del~Corro and
  Gemulla}{2013}]%
        {clausie}
\bibfield{author}{\bibinfo{person}{Luciano Del~Corro} {and}
  \bibinfo{person}{Rainer Gemulla}.} \bibinfo{year}{2013}\natexlab{}.
\newblock \showarticletitle{ClausIE: Clause-based Open Information Extraction}.
  In \bibinfo{booktitle}{{\em Proceedings of the 22Nd International Conference
  on World Wide Web}} {\em (\bibinfo{series}{WWW '13})}.
  \bibinfo{pages}{355--366}.
\newblock


\bibitem[\protect\citeauthoryear{Etzioni, Fader, Christensen, Soderland, and
  Mausam}{Etzioni et~al\mbox{.}}{2011}]%
        {reverb}
\bibfield{author}{\bibinfo{person}{Oren Etzioni}, \bibinfo{person}{Anthony
  Fader}, \bibinfo{person}{Janara Christensen}, \bibinfo{person}{Stephen
  Soderland}, {and} \bibinfo{person}{Mausam Mausam}.}
  \bibinfo{year}{2011}\natexlab{}.
\newblock \showarticletitle{Open Information Extraction: The Second
  Generation}. In \bibinfo{booktitle}{{\em Proceedings of the Twenty-Second
  International Joint Conference on Artificial Intelligence - Volume One}} {\em
  (\bibinfo{series}{IJCAI'11})}. \bibinfo{publisher}{AAAI Press},
  \bibinfo{pages}{3--10}.
\newblock
\showISBNx{978-1-57735-513-7}


\bibitem[\protect\citeauthoryear{F{\"a}rber, Rettinger, and Harth}{F{\"a}rber
  et~al\mbox{.}}{2016}]%
        {farber}
\bibfield{author}{\bibinfo{person}{Michael F{\"a}rber}, \bibinfo{person}{Achim
  Rettinger}, {and} \bibinfo{person}{Andreas Harth}.}
  \bibinfo{year}{2016}\natexlab{}.
\newblock \bibinfo{booktitle}{{\em Towards Monitoring of Novel Statements in
  the News}}.
\newblock \bibinfo{publisher}{Springer International Publishing},
  \bibinfo{address}{Cham}, \bibinfo{pages}{285--299}.
\newblock


\bibitem[\protect\citeauthoryear{Ferrucci, Brown, Chu-Carroll, Fan, Gondek,
  Kalyanpur, Lally, Murdock, Nyberg, Prager, Schlaefer, and Welty}{Ferrucci
  et~al\mbox{.}}{2010}]%
        {watson}
\bibfield{author}{\bibinfo{person}{David Ferrucci}, \bibinfo{person}{Eric
  Brown}, \bibinfo{person}{Jennifer Chu-Carroll}, \bibinfo{person}{James Fan},
  \bibinfo{person}{David Gondek}, \bibinfo{person}{Aditya~A. Kalyanpur},
  \bibinfo{person}{Adam Lally}, \bibinfo{person}{J.~William Murdock},
  \bibinfo{person}{Eric Nyberg}, \bibinfo{person}{John Prager},
  \bibinfo{person}{Nico Schlaefer}, {and} \bibinfo{person}{Chris Welty}.}
  \bibinfo{year}{2010}\natexlab{}.
\newblock \showarticletitle{{The AI Behind Watson -- The Technical Article}}.
\newblock \bibinfo{journal}{{\em The AI Magazine\/}} (\bibinfo{year}{2010}).
\newblock
\showURL{%
\url{http://www.aaai.org/Magazine/Watson/watson.php}}


\bibitem[\protect\citeauthoryear{Fleischhacker, V{\"o}lker, and
  Stuckenschmidt}{Fleischhacker et~al\mbox{.}}{2012}]%
        {propaxms}
\bibfield{author}{\bibinfo{person}{Daniel Fleischhacker},
  \bibinfo{person}{Johanna V{\"o}lker}, {and} \bibinfo{person}{Heiner
  Stuckenschmidt}.} \bibinfo{year}{2012}\natexlab{}.
\newblock \bibinfo{booktitle}{{\em Mining RDF Data for Property Axioms}}.
\newblock \bibinfo{publisher}{Springer Berlin Heidelberg},
  \bibinfo{address}{Berlin, Heidelberg}.
\newblock


\bibitem[\protect\citeauthoryear{Frakes and Baeza-Yates}{Frakes and
  Baeza-Yates}{1992}]%
        {singlepassclustering}
\bibfield{editor}{\bibinfo{person}{William~B. Frakes} {and}
  \bibinfo{person}{Ricardo Baeza-Yates}} (Eds.).
  \bibinfo{year}{1992}\natexlab{}.
\newblock \bibinfo{booktitle}{{\em Information Retrieval: Data Structures and
  Algorithms}}.
\newblock \bibinfo{publisher}{Prentice-Hall, Inc.}, \bibinfo{address}{Upper
  Saddle River, NJ, USA}.
\newblock
\showISBNx{0-13-463837-9}


\bibitem[\protect\citeauthoryear{Gunaratna, Lalithsena, and Sheth}{Gunaratna
  et~al\mbox{.}}{2014}]%
        {all3}
\bibfield{author}{\bibinfo{person}{Kalpa Gunaratna}, \bibinfo{person}{Sarasi
  Lalithsena}, {and} \bibinfo{person}{Amit Sheth}.}
  \bibinfo{year}{2014}\natexlab{}.
\newblock \showarticletitle{Alignment and dataset identification of linked data
  in Semantic Web}.
\newblock \bibinfo{journal}{{\em Wiley Interdisciplinary Reviews: Data Mining
  and Knowledge Discovery\/}} \bibinfo{volume}{4}, \bibinfo{number}{2}
  (\bibinfo{year}{2014}), \bibinfo{pages}{139--151}.
\newblock
\showISSN{1942-4795}


\bibitem[\protect\citeauthoryear{Jain, Hitzler, Verma, Yeh, and Sheth}{Jain
  et~al\mbox{.}}{2012}]%
        {plato}
\bibfield{author}{\bibinfo{person}{Prateek Jain}, \bibinfo{person}{Pascal
  Hitzler}, \bibinfo{person}{Kunal Verma}, \bibinfo{person}{Peter~Z. Yeh},
  {and} \bibinfo{person}{Amit~P. Sheth}.} \bibinfo{year}{2012}\natexlab{}.
\newblock \showarticletitle{Moving Beyond SameAs with PLATO: Partonomy
  Detection for Linked Data}. In \bibinfo{booktitle}{{\em Proceedings of the
  23rd ACM Conference on Hypertext and Social Media}} {\em (\bibinfo{series}{HT
  '12})}. \bibinfo{publisher}{ACM}, \bibinfo{address}{New York, NY, USA},
  \bibinfo{pages}{33--42}.
\newblock
\showISBNx{978-1-4503-1335-3}


\bibitem[\protect\citeauthoryear{Krause, Li, Uszkoreit, and Xu}{Krause
  et~al\mbox{.}}{2012a}]%
        {distantlargescale}
\bibfield{author}{\bibinfo{person}{Sebastian Krause}, \bibinfo{person}{Hong
  Li}, \bibinfo{person}{Hans Uszkoreit}, {and} \bibinfo{person}{Feiyu Xu}.}
  \bibinfo{year}{2012}\natexlab{a}.
\newblock \showarticletitle{Large-Scale Learning of Relation-Extraction Rules
  with Distant Supervision from the Web}.
\newblock In \bibinfo{booktitle}{{\em The Semantic Web - ISWC 2012}}.
  \bibinfo{series}{Lecture Notes in Computer Science},
  Vol.~\bibinfo{volume}{7649}. \bibinfo{publisher}{Springer Berlin Heidelberg},
  \bibinfo{pages}{263--278}.
\newblock
\showISBNx{978-3-642-35175-4}


\bibitem[\protect\citeauthoryear{Krause, Li, Uszkoreit, and Xu}{Krause
  et~al\mbox{.}}{2012b}]%
        {dist}
\bibfield{author}{\bibinfo{person}{Sebastian Krause}, \bibinfo{person}{Hong
  Li}, \bibinfo{person}{Hans Uszkoreit}, {and} \bibinfo{person}{Feiyu Xu}.}
  \bibinfo{year}{2012}\natexlab{b}.
\newblock \showarticletitle{Large-Scale Learning of Relation-Extraction Rules
  with Distant Supervision from the Web}.
\newblock In \bibinfo{booktitle}{{\em The Semantic Web - ISWC 2012}}.
  \bibinfo{series}{Lecture Notes in Computer Science},
  Vol.~\bibinfo{volume}{7649}. \bibinfo{publisher}{Springer Berlin Heidelberg},
  \bibinfo{pages}{263--278}.
\newblock
\showISBNx{978-3-642-35175-4}


\bibitem[\protect\citeauthoryear{Lehmann, Isele, Jakob, Jentzsch, Kontokostas,
  Mendes, Hellmann, Morsey, van Kleef, Auer, and Bizer}{Lehmann
  et~al\mbox{.}}{2015}]%
        {dbpedia}
\bibfield{author}{\bibinfo{person}{Jens Lehmann}, \bibinfo{person}{Robert
  Isele}, \bibinfo{person}{Max Jakob}, \bibinfo{person}{Anja Jentzsch},
  \bibinfo{person}{Dimitris Kontokostas}, \bibinfo{person}{Pablo~N. Mendes},
  \bibinfo{person}{Sebastian Hellmann}, \bibinfo{person}{Mohamed Morsey},
  \bibinfo{person}{Patrick van Kleef}, \bibinfo{person}{Sören Auer}, {and}
  \bibinfo{person}{Christian Bizer}.} \bibinfo{year}{2015}\natexlab{}.
\newblock \showarticletitle{DBpedia - A large-scale, multilingual knowledge
  base extracted from Wikipedia.}
\newblock \bibinfo{journal}{{\em Semantic Web\/}}  \bibinfo{volume}{6}
  (\bibinfo{year}{2015}), \bibinfo{pages}{167--195}.
\newblock


\bibitem[\protect\citeauthoryear{Limaye, Sarawagi, and Chakrabarti}{Limaye
  et~al\mbox{.}}{2010}]%
        {iitb}
\bibfield{author}{\bibinfo{person}{Girija Limaye}, \bibinfo{person}{Sunita
  Sarawagi}, {and} \bibinfo{person}{Soumen Chakrabarti}.}
  \bibinfo{year}{2010}\natexlab{}.
\newblock \showarticletitle{Annotating and Searching Web Tables Using Entities,
  Types and Relationships}.
\newblock \bibinfo{journal}{{\em Proc. VLDB Endow.\/}} \bibinfo{volume}{3},
  \bibinfo{number}{1-2} (\bibinfo{date}{Sept.} \bibinfo{year}{2010}),
  \bibinfo{pages}{1338--1347}.
\newblock
\showISSN{2150-8097}


\bibitem[\protect\citeauthoryear{Mahdisoltani, Biega, and
  Suchanek}{Mahdisoltani et~al\mbox{.}}{2015}]%
        {yago3}
\bibfield{author}{\bibinfo{person}{Farzaneh Mahdisoltani},
  \bibinfo{person}{Joanna Biega}, {and} \bibinfo{person}{Fabian~M. Suchanek}.}
  \bibinfo{year}{2015}\natexlab{}.
\newblock \showarticletitle{{YAGO3:} {A} Knowledge Base from Multilingual
  Wikipedias}. In \bibinfo{booktitle}{{\em {CIDR} 2015, Seventh Biennial
  Conference on Innovative Data Systems Research, Asilomar, CA, USA, January
  4-7, 2015, Online Proceedings}}.
\newblock


\bibitem[\protect\citeauthoryear{Mausam, Schmitz, Bart, Soderland, and
  Etzioni}{Mausam et~al\mbox{.}}{2012}]%
        {ollie}
\bibfield{author}{\bibinfo{person}{Mausam}, \bibinfo{person}{Michael Schmitz},
  \bibinfo{person}{Robert Bart}, \bibinfo{person}{Stephen Soderland}, {and}
  \bibinfo{person}{Oren Etzioni}.} \bibinfo{year}{2012}\natexlab{}.
\newblock \showarticletitle{Open Language Learning for Information Extraction}.
  In \bibinfo{booktitle}{{\em Proceedings of the 2012 Joint Conference on
  Empirical Methods in Natural Language Processing and Computational Natural
  Language Learning}} {\em (\bibinfo{series}{EMNLP-CoNLL '12})}.
  \bibinfo{pages}{523--534}.
\newblock


\bibitem[\protect\citeauthoryear{Mihalcea, Corley, and Strapparava}{Mihalcea
  et~al\mbox{.}}{2006}]%
        {mihalcea}
\bibfield{author}{\bibinfo{person}{Rada Mihalcea}, \bibinfo{person}{Courtney
  Corley}, {and} \bibinfo{person}{Carlo Strapparava}.}
  \bibinfo{year}{2006}\natexlab{}.
\newblock \showarticletitle{Corpus-based and Knowledge-based Measures of Text
  Semantic Similarity}. In \bibinfo{booktitle}{{\em Proceedings of the 21st
  National Conference on Artificial Intelligence - Volume 1}} {\em
  (\bibinfo{series}{AAAI'06})}. \bibinfo{publisher}{AAAI Press},
  \bibinfo{pages}{775--780}.
\newblock


\bibitem[\protect\citeauthoryear{Mikolov, Sutskever, Chen, Corrado, and
  Dean}{Mikolov et~al\mbox{.}}{2013}]%
        {word2vec}
\bibfield{author}{\bibinfo{person}{Tomas Mikolov}, \bibinfo{person}{Ilya
  Sutskever}, \bibinfo{person}{Kai Chen}, \bibinfo{person}{Greg~S Corrado},
  {and} \bibinfo{person}{Jeff Dean}.} \bibinfo{year}{2013}\natexlab{}.
\newblock \showarticletitle{Distributed Representations of Words and Phrases
  and their Compositionality}.
\newblock In \bibinfo{booktitle}{{\em Advances in Neural Information Processing
  Systems 26}}, \bibfield{editor}{\bibinfo{person}{C.~J.~C. Burges},
  \bibinfo{person}{L.~Bottou}, \bibinfo{person}{M.~Welling},
  \bibinfo{person}{Z.~Ghahramani}, {and} \bibinfo{person}{K.~Q. Weinberger}}
  (Eds.). \bibinfo{pages}{3111--3119}.
\newblock


\bibitem[\protect\citeauthoryear{Mintz, Bills, Snow, and Jurafsky}{Mintz
  et~al\mbox{.}}{2009}]%
        {mintz}
\bibfield{author}{\bibinfo{person}{Mike Mintz}, \bibinfo{person}{Steven Bills},
  \bibinfo{person}{Rion Snow}, {and} \bibinfo{person}{Dan Jurafsky}.}
  \bibinfo{year}{2009}\natexlab{}.
\newblock \showarticletitle{Distant Supervision for Relation Extraction Without
  Labeled Data}. In \bibinfo{booktitle}{{\em Proceedings of the Joint
  Conference of the 47th Annual Meeting of the ACL and the 4th International
  Joint Conference on Natural Language Processing of the AFNLP: Volume 2 -
  Volume 2}} {\em (\bibinfo{series}{ACL '09})}. \bibinfo{publisher}{Association
  for Computational Linguistics}, \bibinfo{address}{Stroudsburg, PA, USA},
  \bibinfo{pages}{1003--1011}.
\newblock
\showISBNx{978-1-932432-46-6}


\bibitem[\protect\citeauthoryear{Mohamed, Hruschka, and Mitchell}{Mohamed
  et~al\mbox{.}}{2011}]%
        {ontext}
\bibfield{author}{\bibinfo{person}{Thahir~P. Mohamed},
  \bibinfo{person}{Estevam~R. Hruschka, Jr.}, {and} \bibinfo{person}{Tom~M.
  Mitchell}.} \bibinfo{year}{2011}\natexlab{}.
\newblock \showarticletitle{Discovering Relations Between Noun Categories}. In
  \bibinfo{booktitle}{{\em Proceedings of the Conference on Empirical Methods
  in Natural Language Processing}} {\em (\bibinfo{series}{EMNLP '11})}.
  \bibinfo{pages}{1447--1455}.
\newblock
\showISBNx{978-1-937284-11-4}


\bibitem[\protect\citeauthoryear{Mu\~{n}oz, Hogan, and Mileo}{Mu\~{n}oz
  et~al\mbox{.}}{2013}]%
        {triplifying2}
\bibfield{author}{\bibinfo{person}{Emir Mu\~{n}oz}, \bibinfo{person}{Aidan
  Hogan}, {and} \bibinfo{person}{Alessandra Mileo}.}
  \bibinfo{year}{2013}\natexlab{}.
\newblock \showarticletitle{Triplifying Wikipedia's Tables.}. In
  \bibinfo{booktitle}{{\em LD4IE@ISWC}} {\em (\bibinfo{series}{CEUR Workshop
  Proceedings})}, \bibfield{editor}{\bibinfo{person}{Anna~Lisa Gentile},
  \bibinfo{person}{Ziqi Zhang}, \bibinfo{person}{Claudia d'Amato}, {and}
  \bibinfo{person}{Heiko Paulheim}} (Eds.), Vol.~\bibinfo{volume}{1057}.
  \bibinfo{publisher}{CEUR-WS.org}.
\newblock


\bibitem[\protect\citeauthoryear{Mu\~{n}oz, Hogan, and Mileo}{Mu\~{n}oz
  et~al\mbox{.}}{2014}]%
        {triplifying1}
\bibfield{author}{\bibinfo{person}{Emir Mu\~{n}oz}, \bibinfo{person}{Aidan
  Hogan}, {and} \bibinfo{person}{Alessandra Mileo}.}
  \bibinfo{year}{2014}\natexlab{}.
\newblock \showarticletitle{Using Linked Data to Mine RDF from Wikipedia's
  Tables}. In \bibinfo{booktitle}{{\em Proceedings of the 7th ACM International
  Conference on Web Search and Data Mining}} {\em (\bibinfo{series}{WSDM
  '14})}. \bibinfo{publisher}{ACM}, \bibinfo{address}{New York, NY, USA},
  \bibinfo{pages}{533--542}.
\newblock
\showISBNx{978-1-4503-2351-2}


\bibitem[\protect\citeauthoryear{Mulwad}{Mulwad}{2010}]%
        {t2ld2}
\bibfield{author}{\bibinfo{person}{Varish Mulwad}.}
  \bibinfo{year}{2010}\natexlab{}.
\newblock {\em \bibinfo{title}{{T2LD - An automatic framework for extracting,
  interpreting and representing tables as Linked Data}}}.
\newblock \bibinfo{thesistype}{Master's\ thesis}.
\newblock


\bibitem[\protect\citeauthoryear{Mulwad, Finin, Syed, and Joshi}{Mulwad
  et~al\mbox{.}}{2010a}]%
        {t2ld1}
\bibfield{author}{\bibinfo{person}{Varish Mulwad}, \bibinfo{person}{Tim Finin},
  \bibinfo{person}{Zareen Syed}, {and} \bibinfo{person}{Anupam Joshi}.}
  \bibinfo{year}{2010}\natexlab{a}.
\newblock \showarticletitle{{T2LD:} Interpreting and Representing Tables as
  Linked Data}. In \bibinfo{booktitle}{{\em Proceedings of the {ISWC} 2010
  Posters {\&} Demonstrations Track: Collected Abstracts, Shanghai, China,
  November 9, 2010}}.
\newblock


\bibitem[\protect\citeauthoryear{Mulwad, Finin, Syed, and Joshi}{Mulwad
  et~al\mbox{.}}{2010b}]%
        {mulwad2}
\bibfield{author}{\bibinfo{person}{Varish Mulwad}, \bibinfo{person}{Tim Finin},
  \bibinfo{person}{Zareen Syed}, {and} \bibinfo{person}{Anupam Joshi}.}
  \bibinfo{year}{2010}\natexlab{b}.
\newblock \showarticletitle{Using Linked Data to Interpret Tables}. In
  \bibinfo{booktitle}{{\em Proceedings of the First International Workshop on
  Consuming Linked Data, Shanghai, China, November 8, 2010}}.
\newblock


\bibitem[\protect\citeauthoryear{Navarro}{Navarro}{2016}]%
        {disslfn}
\bibfield{author}{\bibinfo{person}{Lucas~Fonseca Navarro}.}
  \bibinfo{year}{2016}\natexlab{}.
\newblock {\em \bibinfo{title}{Mining Ontologies to Extract Implicit
  Knowledge}}.
\newblock \bibinfo{thesistype}{Ph.D. Dissertation}. \bibinfo{school}{Federal
  University of Sao Carlos}.
\newblock


\bibitem[\protect\citeauthoryear{Nguyen and Moschitti}{Nguyen and
  Moschitti}{2011}]%
        {yagowikipedia}
\bibfield{author}{\bibinfo{person}{Truc-Vien~T. Nguyen} {and}
  \bibinfo{person}{Alessandro Moschitti}.} \bibinfo{year}{2011}\natexlab{}.
\newblock \showarticletitle{End-to-end Relation Extraction Using Distant
  Supervision from External Semantic Repositories}. In \bibinfo{booktitle}{{\em
  Proceedings of the 49th Annual Meeting of the Association for Computational
  Linguistics: Human Language Technologies: Short Papers - Volume 2}} {\em
  (\bibinfo{series}{HLT '11})}. \bibinfo{publisher}{Association for
  Computational Linguistics}, \bibinfo{address}{Stroudsburg, PA, USA},
  \bibinfo{pages}{277--282}.
\newblock
\showISBNx{978-1-932432-88-6}


\bibitem[\protect\citeauthoryear{Paulheim}{Paulheim}{2017}]%
        {swjsurvey}
\bibfield{author}{\bibinfo{person}{Heiko Paulheim}.}
  \bibinfo{year}{2017}\natexlab{}.
\newblock \showarticletitle{Knowledge graph refinement: {A} survey of
  approaches and evaluation methods}.
\newblock \bibinfo{journal}{{\em Semantic Web\/}} \bibinfo{volume}{8},
  \bibinfo{number}{3} (\bibinfo{year}{2017}), \bibinfo{pages}{489--508}.
\newblock


\bibitem[\protect\citeauthoryear{Rajaraman and Ullman}{Rajaraman and
  Ullman}{2011}]%
        {mmdbook}
\bibfield{author}{\bibinfo{person}{Anand Rajaraman} {and}
  \bibinfo{person}{Jeffrey~David Ullman}.} \bibinfo{year}{2011}\natexlab{}.
\newblock \bibinfo{booktitle}{{\em Mining of Massive Datasets}}.
\newblock \bibinfo{publisher}{Cambridge University Press},
  \bibinfo{address}{New York, NY, USA}.
\newblock
\showISBNx{1107015359, 9781107015357}


\bibitem[\protect\citeauthoryear{Ritze, Lehmberg, and Bizer}{Ritze
  et~al\mbox{.}}{2015}]%
        {T2Kmatch}
\bibfield{author}{\bibinfo{person}{Dominique Ritze}, \bibinfo{person}{Oliver
  Lehmberg}, {and} \bibinfo{person}{Christian Bizer}.}
  \bibinfo{year}{2015}\natexlab{}.
\newblock \showarticletitle{Matching HTML Tables to DBpedia}. In
  \bibinfo{booktitle}{{\em Proceedings of the 5th International Conference on
  Web Intelligence, Mining and Semantics}} {\em (\bibinfo{series}{WIMS '15})}.
  \bibinfo{publisher}{ACM}, Article \bibinfo{articleno}{10},
  \bibinfo{numpages}{6}~pages.
\newblock


\bibitem[\protect\citeauthoryear{Suchanek, Kasneci, and Weikum}{Suchanek
  et~al\mbox{.}}{2007}]%
        {yago}
\bibfield{author}{\bibinfo{person}{Fabian~M. Suchanek},
  \bibinfo{person}{Gjergji Kasneci}, {and} \bibinfo{person}{Gerhard Weikum}.}
  \bibinfo{year}{2007}\natexlab{}.
\newblock \showarticletitle{Yago: A Core of Semantic Knowledge}. In
  \bibinfo{booktitle}{{\em Proceedings of the 16th International Conference on
  World Wide Web}} {\em (\bibinfo{series}{WWW '07})}.
  \bibinfo{pages}{697--706}.
\newblock
\showISBNx{978-1-59593-654-7}


\bibitem[\protect\citeauthoryear{Suchanek, Sozio, and Weikum}{Suchanek
  et~al\mbox{.}}{2009}]%
        {sofie}
\bibfield{author}{\bibinfo{person}{Fabian~M. Suchanek}, \bibinfo{person}{Mauro
  Sozio}, {and} \bibinfo{person}{Gerhard Weikum}.}
  \bibinfo{year}{2009}\natexlab{}.
\newblock \showarticletitle{SOFIE: A Self-organizing Framework for Information
  Extraction}. In \bibinfo{booktitle}{{\em Proceedings of the 18th
  International Conference on World Wide Web}} {\em (\bibinfo{series}{WWW
  '09})}. \bibinfo{publisher}{ACM}, \bibinfo{address}{New York, NY, USA},
  \bibinfo{pages}{631--640}.
\newblock
\showISBNx{978-1-60558-487-4}


\bibitem[\protect\citeauthoryear{Syed, Finin, Mulwad, and Joshi}{Syed
  et~al\mbox{.}}{2010}]%
        {mulwad1}
\bibfield{author}{\bibinfo{person}{Zareen Syed}, \bibinfo{person}{Tim Finin},
  \bibinfo{person}{Varish Mulwad}, {and} \bibinfo{person}{Anupam Joshi}.}
  \bibinfo{year}{2010}\natexlab{}.
\newblock \showarticletitle{A.: Exploiting a Web of Semantic Data for
  Interpreting Tables}. In \bibinfo{booktitle}{{\em In: Proceedings of the
  Second Web Science Conference.}}
\newblock


\bibitem[\protect\citeauthoryear{T\"{o}pper, Knuth, and Sack}{T\"{o}pper
  et~al\mbox{.}}{2012}]%
        {topper}
\bibfield{author}{\bibinfo{person}{Gerald T\"{o}pper}, \bibinfo{person}{Magnus
  Knuth}, {and} \bibinfo{person}{Harald Sack}.}
  \bibinfo{year}{2012}\natexlab{}.
\newblock \showarticletitle{DBpedia Ontology Enrichment for Inconsistency
  Detection}. In \bibinfo{booktitle}{{\em Proceedings of the 8th International
  Conference on Semantic Systems}} {\em (\bibinfo{series}{I-SEMANTICS '12})}.
  \bibinfo{publisher}{ACM}, \bibinfo{address}{New York, NY, USA},
  \bibinfo{pages}{33--40}.
\newblock
\showISBNx{978-1-4503-1112-0}


\end{thebibliography}

\end{document}